%% file: DMbinaries.v6.tex
\documentclass[11pt,a4paper]{article}
\usepackage{jcappub}
\pdfoutput=1

\usepackage{amsfonts}
\usepackage{amsmath}
\usepackage{amssymb}
\usepackage{aas_macros}
\usepackage{graphicx}
\usepackage{color}
\usepackage{float}
\usepackage{braket}
\usepackage{mathrsfs}
\allowdisplaybreaks[1]

\input{defs.tex}

\newcommand{\ham}{\mathscr{H}}
\newcommand{\eps}{\epsilon}
\newcommand{\res}{\textrm{res}}
\newcommand{\ak}{a_*}
\newcommand{\ek}{e_*}
\newcommand{\gk}{\gamma_k}

\newcommand{\abs}[1]{\left\vert#1\right\vert}

\definecolor{RedWine}{rgb}{0.743,0,0}
\definecolor{RoyalBlue}{rgb}{0.25,.41,.88}
\definecolor{ForestGreen}{rgb}{.13,.54,.13}

\newcommand{\be}{\begin{equation}}
\newcommand{\ee}{\end{equation}}
\title{Axion resonances in binary pulsar systems}
\author{Mor Rozner,}
\affiliation{Physics department, Technion, Haifa 3200003, Israel}
\author{Evgeni Grishin,}
\author{Yonadav Barry Ginat,}
\author{Andrei P. Igoshev and}
\author{Vincent Desjacques}

\emailAdd{morozner@campus.technion.ac.il}
\emailAdd{eugeneg@campus.technion.ac.il}
\emailAdd{dvince@physics.technion.ac.il}

\abstract{We investigate the extent to which resonances between an oscillating background of ultra-light axion and a binary Keplerian system can affect the motion of the latter. These resonances lead to perturbations in the instantaneous time-of-arrivals, and to secular variations in the period of the binary. While the secular changes at exact resonance have recently been explored, the instantaneous effects have been overlooked. In this paper, we examine the latter using N-body simulations including the external oscillatory forcing induced by the axion background. While the secular effects are restricted to a narrow width near the resonance, the instantaneous changes, albeit strongest close to resonances, are apparent for wide range of configurations. We compute the signal-to-noise ratio (SNR) as a function of semi-major axis for a detection of axion oscillations through the R\o mer delay. The latter can be extracted from the time-of-arrivals of binary pulsars. The SNR broadly increases with increasing binary eccentricity as expected from secular expectations. However, we find that it differs significantly from the scaling $a^{5/2}$ around the lowest orders of resonance. Future observations could probe these effects away from resonances and, therefore, constrain a much broader range of axion masses provided that binary pulsar systems are found near the central region of our Galaxy, and that the time-or-arrival measurement accuracy reaches $\lesssim 10$ ns.
 }
\begin{document}
\maketitle

\date{\today}
\label{firstpage}

 \section{Introduction}

 The nature of dark matter is one of the greatest puzzles in physics. Since its existence was conjectured by Zwicky \cite{Zwicky:1933} from virial equilibrium considerations in galaxy clusters, a plethora of models have been proposed, ranging from weakly interacting massive particles (WIMPs) \cite{lee/weinberg:1977,steigman/turner:1984} to modified gravity theories such as MOND \cite{Milgrom1983}, but many of them are now severely constrained by astrophysical/cosmological or particle physics experiments. Another possibility is that dark matter is a Bose-Einstein condensate of light bosons such as axions \cite[see][for early work]{Baldeschi:1983mq,Sin:1992bg,hu/etal:2000}. Ultra-light axions with a mass $m_a\gtrsim 10^{-22}\ {\rm eV}$ much lighter than that of QCD axions could help resolve some of the small-scale problems of standard cold dark matter (CDM) such as the core-cusp issue etc. \cite[see, e.g.][for recent comprehensive discussions]{Marsh,TremaineWitten, LopezNacir:2018epg}. However, large Compton wavelengths leading to a significant suppression of the small-scale power spectrum are incompatible with Lyman-$\alpha$ forest measurements, which set a lower bound of $m_a\gtrsim 10^{-21}\rm\ eV$ (95\% C.L.) \cite{Irsic:2017yje,Armengaud:2017nkf}.

Pulsar timing offers another avenue to probe the existence of coherent oscillations induced by ultra-light scalar fields \cite{Khmelnitsky:2013,Blas,FollowupToBlas}. Millisecond radio pulsars (MSPs, for review see \cite{2008LRR....11....8L}) are known as the most stable clocks in the universe on timescales longer than a few spin periods of neutron star. It is, therefore, not surprising that double neutron star systems and neutron star - white dwarfs systems are actively used to study gravity (for review see \cite{2014LRR....17....4W}). In particular, the first indirect detection of gravitational waves emission  \citep{2005ASPC..328...25W,2010ApJ...722.1030W} was based on detail measurements of the orbital period decrease in a binary pulsar. 
Another example are tests of the equivalence principle using a triple containing a MSP and two white dwarfs \citep{2018Natur.559...73A}.
 The central parsec of the Milky Way could contain up to 10000 MSPs and a part of them will be discovered with the next generation radio telescopes such as SKA \citep{2017MNRAS.471..730R}.

When a binary pulsar or MSP with a white dwarf companion orbits in a background of ultra-light axions, the dynamics of the binary is affected by the axion oscillations especially if the axion mass $m_a$ is close to integer multiples of half the orbital frequency $\Omega_c$. When there is such a resonance, axion oscillations induce a secular perturbation to the orbital period which might be observable with pulsar timing experiments \cite{Blas, newBlas}.

In this paper, we investigate the instantaneous perturbation rather than the secular effect using direct N-body simulations, in which the axion background is treated as a small perturbation to the binary's Keplerian solution. Our analysis is, therefore, also valid away from resonances, which are restricted to a very narrow range of orbital frequencies (or semi-major axes).

The paper is organized as follows. In section \S\ref{sec:setup}, we provide a general overview of ultra-light axion dark matter and the resonance phenomenon. In \ref{subsec:osc pert}, we discuss the  perturbations of a Newtonian binary, while \ref{sec:secular} and \ref{sec:resonance} focus on the  secular change of the orbital period, and  the instantaneous effect of resonances, respectively. In section \S\ref{sec:nbody}, we introduce the N-body simulations and present the expected signals.
In section \S\ref{sec: prospects} we discuss the prospects for a detection of axion oscillations through the measurement of the R\o mer delay in pulsar timing experiments.
We summarize our conclusions in section \S\ref{sec:conclusions}.

 \section{Theoretical background}\label{sec:setup}

Ultra-light axion fields with mass $m_a$ in the range $10^{-22} - 10^{-20}\  \rm eV$  oscillate on characteristic timescales of a few hours to a few days, which are of the same order as the orbital periods $T$ of typical binary pulsar systems. More precisely, when the orbital frequency $\Omega_c=2\pi/T$ multiplied by an integer equals twice the axion mass $m_a$,
\begin{equation} \label{eq:resonancecondition}
\mbox{{\it resonance condition}}:\quad \Omega_c k = 2m_a,\quad k \in \mathbb{N}
\end{equation}
there is a resonance between the motion of the binary and the axion background. Let us first revisit the analysis of \cite{Khmelnitsky:2013,Blas,FollowupToBlas,newBlas}.

Note that, unless stated otherwise, we use natural units $c=\hbar=k_B=1$ throughout.

 \subsection{Oscillatory Perturbations of a Newtonian Binary}
 \label{subsec:osc pert}

 The evolution of the ultra-light axion field is described by the Klein-Gordon equation, which solution is of the form \cite[e.g.][for a review]{Marsh}
\begin{equation}
\phi(\mathbf{x}, t)=\psi(\mathbf{x})\cos\big(m_a t+\varphi({\bf x})\big)
\end{equation}
 where $\psi$ is a slowly varying complex amplitude (we thus ignore its time dependence since it is much longer than the timescales involved in our problem), $m_a$ is the axion mass and $\varphi({\bf x})$ is a position-dependent phase.

 The energy density and pressure of the axion field can be extracted from the stress-energy tensor. They can be separated into two components: a time-independent piece, and an oscillating term arising from the coherent oscillations of the axion field. While the dominant contribution to the energy density slowly varies with time and scales like $\propto \psi^2({\bf x})\propto a^{-3}$, the dominant contribution to the pressure oscillates on a timescale $\propto m_a^{-1}$ according to
  \begin{equation}
 P_{\rm DM}({\bf x},t)\supset -\frac{1}{2}m_a^2 \psi^2({\bf x})\cos(2m_at+ 2\varphi) \;,
 \end{equation}
 that is, a characteristic frequency $\omega_{a}=2m_a$ twice the axion mass.

 In order to investigate the effects of this perturbation, we consider a perturbed metric in the Newtonian gauge
 \begin{align}
 ds^2=(1+2\Psi)dt^2+(1-2\Phi)\delta_{ij}d x^i dx^j \;,
 \end{align}
 in which $\Psi$ is the Newtonian gravitational potential sourced by the dark matter distribution. The spatial part of Einstein's equations shows that the time-independent contributions to the scalar potentials satisfy $\Psi=\Phi$ \cite[the time-independent anisotropic stress induced by axion field is negligible, see e.g.][]{Vicncent&Mor}, while the time-dependent pressure generates a time-dependent gravitational potential \cite{Khmelnitsky:2013}
 \begin{align}
 \Psi(\mathbf{x},t) \supset\Psi_{2c}({\bf x})\cos(\omega_at+2\varphi)
 \end{align}
 with an amplitude
 \begin{equation}
 \Psi_\text{2c}({\bf x}) = \frac{\pi G}{m_a^2}\rho_\text{DM}({\bf x}) \;,\quad
 \rho_\text{DM}({\bf x})=\frac{1}{2}m_a^2 \psi^2({\bf x})
 \end{equation}
The effect is proportional to the axion density and, therefore, is largest at the center of the axion solitonic cores where the density reaches $\rho_{\rm DM}({\bf x})\approx 10^2 (m_a/10^{-22}\ \rm eV)^2\ \rm M_{\odot} \ pc^{-3}$ \cite{Chavanis:2011zi} for present-day Milky-Way size halos.

 \subsection{Secular change of the orbital period}
 \label{sec:secular}
 
 Consider the motion of a binary with separation $a$, total mass $M$ and reduced mass $\mu$ in a background of ultra-light axions. In the local Fermi frame attached to the binary center-of-mass, the perturbative force induced by the oscillatory background of axions is \cite{Blas}
 \begin{equation}\label{eqn:axion force}
 \mathbf{F} = -4\pi G\rho_{\rm DM}\mu\cos(\omega_a t + \varphi)\mathbf{r}
 \end{equation}
 where $\omega_a=2m_a$. This adds a time-dependent piece to the Hamiltonian given by
 \begin{equation}
 \ham_{\textrm{pert}} = 2\pi G\rho_{\rm DM}\mu\cos(\omega_a t + \varphi)r^2.
 \end{equation}
The strength of the perturbation is quantified by the ratio of the typical strength of the perturbative and Keplerian Hamiltonians $\ham_{\rm pert}^0$ and $\ham_K^0$:
\begin{equation}\label{eqn:definition of epsilon}
     \epsilon \equiv \frac{\ham_{\rm pert}^0}{\ham_{K}^0} = \frac{2\pi G \rho_{\rm DM} \mu \ak^2}{G M \mu /\ak} = \frac{2\pi\rho_{\rm DM}\ak^3}{M},
 \end{equation}
where $\ak$ is the semi-major axis (SMA) at the $k$-th order of resonance. From Kepler's law, it is given by
 \begin{equation}
    \label{eq: kth resonance}
     \ak = \left(\frac{GM}{\omega_{a}^{2}}\right)^{1/3}k^{2/3} , \ k\in \mathbb N
 \end{equation}
 Unsuprisingly, we have $\epsilon\propto \ak^3$ since the relative amplitude of the axion oscillations depends on dark matter mass enclosed by the orbit.

  We show in Appendix \S\ref{app:resonancebit} that, upon isolating the resonant part of the perturbed Hamiltonian, the change in the orbital period derived by \citep{Blas} through a secular averaging is recovered:
 \begin{equation}
 \dot{T}  = -6\frac{J_k(k\ek)}{k}T^2G\rho_{\rm DM}\sin(\gk + \varphi), \label{eq:blastime}
 \end{equation}
 Here, $J_k$ is the $k$-th Bessel function, whereas $\ek$ and $\gk$ are the corresponding eccentricity and angle evaluated for the resonance condition Eq. (\ref{eq:resonancecondition}).

 The impact of a resonance reveals itself in the long-term evolution of the system, rather than the instantaneous accelerations. Its increased influence on the motion of the system arises due to the extended coherence of the forces. However, we emphasize that the system also accumulates instantaneous changes which, as we shall see below, gather to somewhat different effects as one moves away from resonances.

 \subsection{\textbf{Motion close to resonance}}
 \label{sec:resonance}

To reveal the effect of resonances (see e.g. \cite{Arnold}), we adopt the following strategy: we work in the extended phase-space, so that the collection of points that satisfy the resonance condition ($\omega_a$ being an integer multiple of $\Omega_c$) is an hypersurface of the extended phase space -- the resonant surface. Since we are chiefly interested in points near this surface, we transform out the angle-action variables so that the resonance condition no longer is a condition on a linear combination of the angles, diagonalising it in effect. This results in a new set of angle-action variables, which includes a slowly evolving resonant angle and its conjugate momentum.

More precisely, the extended phase-space is constructed by introducing the canonical (angle-action) variables $(t, -E)$, where the time coordinate $t$ is now an additional angle-like variable, and $-E$ is the corresponding generalized momentum. A new variable $\tau$ is introduced to parametrize the phase space trajectories~\footnote{In analogy with Special Relativity, $\tau$ plays the role of proper time while $t$ is the time coordinate. Note that, in the analytical mechanics literature, both symbols are usually interchanged.}.
Furthermore, the Hamiltonian transforms as $\ham\to\ham_{\textrm{new}} = \ham - E$. We shall henceforth drop the subscript `new' on the extended Hamiltonian to avoid notational encumbrance. The resonant coordinates (say, for the $k$-th resonance) are defined as follows: the resonant angle is $\gk = \omega_a t - k\theta_c$, where $\theta_c$ is the true anomaly, whereas the corresponding resonant action is $J_{\rm res}$. We refer the reader to Appendix \ref{app:resonancebit} for more details.

\begin{figure}[h!]
 \centering
 	\includegraphics[width=10cm]{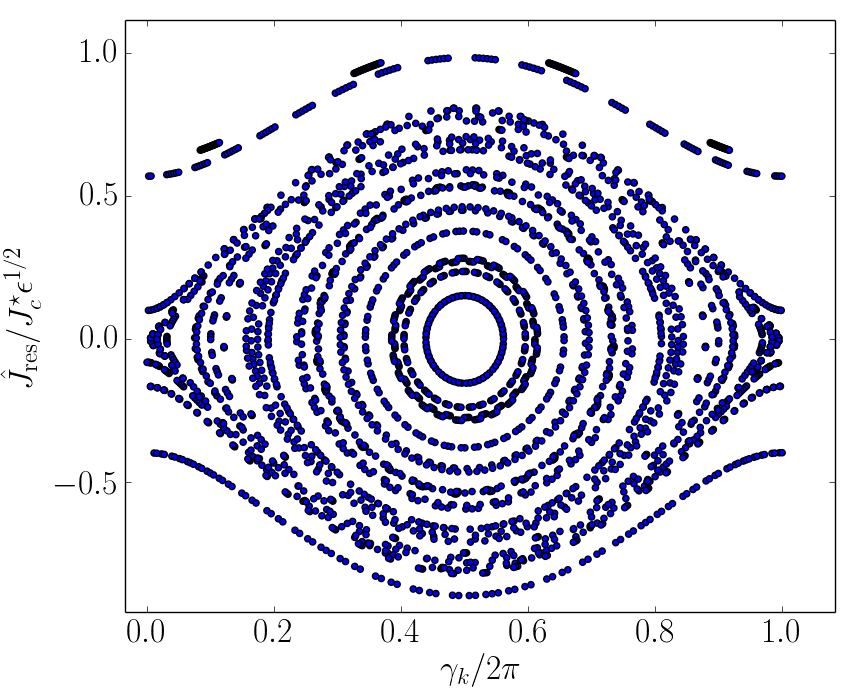}
 	\caption{Phase portrait of $(\hat{J}_{{\rm res}},\gamma_{k})$ obtained upon solving the equations \eqref{eq:pendulumeqs} with the assumption $\varphi=0$.
          We see a fixed stable point at $\gamma_{k}=\pi$, and unstable fixed point at $\gamma_{k}=0$. For $M=M_\odot$, $\mu=M/2$, $k=1$, $a_*=0.2052$ AU, $\epsilon=10^{-13}$ and $e_*=0.5$, the libration timescale is of order $\simeq 2.45\times 10^5$ yr. The width of the separatrix is $\sim\epsilon^{1/2}J_{c}^{\star}$, as expected. The fast oscillations (neglected here) will induce fluctuations around the orbits shown in this figure.
 	}
 	\label{fig0}
 \end{figure}

The resonant coordinates $\gk$ and $J_\text{res}$ do not change when the system is on the resonant surface, while they change slowly near resonance.
Hence, we move on to expand the Hamiltonian about the resonant surface in powers of $\epsilon$, as well as average over the non-resonant angles (which evolve fast). All the non-resonant actions are integrals of motion up to first order in $\eps$, and assume their values at the resonant surface. Furthermore, the non-resonant part of the Hamiltonian oscillates rapidly in the non-resonant -- or fast -- angles, and average out to zero. Therefore,
we conveniently define $J_{\textrm{res}}^*$ as the value of the resonant action $J_\textrm{res}$ at the resonant surface. If its deviation from this surface $\hat{J}_\textrm{res} = J_\textrm{res} - J_{\textrm{res}}^*$ is small (i.e. of order $\mathcal{O}(\sqrt{\epsilon})$) as the system evolves around the resonant surface, we can expand the Hamiltonian in powers of $\hat J_\text{res}$. Retaining terms up to quadratic order, that is, of order $\epsilon$, we arrive at (cf. Appendix \ref{app:resonancebit})
\begin{align}
   \ham(\gk,\hat J_\text{res})&\approx - \frac{G^2M^2\mu^3}{2J_c^{*2}}- E^*   \label{eq:1} \\
   &\quad -\frac{3k^2G^2M^2\mu^3}{J_c^{*4}}\frac{\hat{J}_\res^2}{2}-\frac{2\epsilon}{k^2}J_k(k\ek)\cos(\gk + \varphi)\frac{GM\mu}{\ak} \nonumber \;.
\end{align}
On the right-hand side, the first line is an irrelevant constant while the second line is the Hamiltonian of a simple pendulum, in which
  $\hat{J}_\res$ plays the role of the momentum and $\gk$ the angle subtended relative to the vertical axis.
The equations of motion are 
\begin{align}
  \label{eq:pendulumeqs}
\frac{d\gamma_{k}}{d\tau} & =+\frac{\partial H}{\partial\hat{J}_{{\rm res}}}=-\frac{3k^{2}G^{2}M^{2}}{J_{c}^{\star4}}\mu^{3}\hat{J}_{{\rm res}}=-\frac{3k^{2}}{\mu a_{\star}^{2}}\hat{J}_{{\rm res}}\\
\frac{d\hat{J}_\text{res}}{d\tau} &= -\frac{\partial H}{\partial\gk}=-\frac{2GM\mu\epsilon}{a_{\star}k^{2}}J_{k}(ke_{\star})\sin(\gamma_{k}+\phi) \nonumber \;.
\end{align}
These equations are supplemented by
\begin{equation}
\frac{dt}{d\tau} = +\frac{\partial\ham}{\partial(-E)} = 1 \;,
\end{equation}
which shows that $\tau\equiv t$ without loss of generality. There are two fixed points $(\gk,\hat J_\text{res})=(-\varphi\pm\pi,0)$ and $(-\varphi,0)$, respectively stable and unstable. They correspond to the resonant (phase space) orbits, which arise when i) the resonance condition is satisfied and ii) the oscillatory force induced by the axion condensate is maximally outwards (stable fixed point) or inwards (unstable fixed point) at periapsis.
Overall, $\ham$ generates two distinct classes of orbits: libration (small oscillations about the stable equilibrium) very near the resonance; and circulation or rotation (the pendulum completes whole revolutions) as one moves away from it. A phase space portrait is shown in Fig.\ref{fig0}.

During such periodic motions, which take place on a timescale $\propto 1/\sqrt{\epsilon}$, one can derive the typical variation in $k\hat{J}_\res = J_c^* - J_c$ by considering e.g. the average kinetic energy of a pendulum initially at rest at the unstable equilibrium  point (i.e. at the top). The result is
\begin{equation}
  k\delta(\hat{J}_\res) \sim \left( \frac{8\epsilon J_k(k\ek)}{3k^2}\right)^{1/2} J_c^*.
\end{equation}
The scaling $k\delta(\hat J_\text{res})\propto \sqrt{\epsilon}J_c^*$ is actually valid within the entire region of libration, and inside the region of rotation so long as corrections of order $\mathcal{O}(\hat J_\text{res}^3)$ to \eqref{eq:1} are negligible.

Recall that $J_c$ is associated with the semi-major axis of the binary, and thus determines its period according to Kepler's third law. Using this relation between $J_c$ and $T$, we observe that the period changes by
\begin{equation}\label{eqn:resonant change}
 \frac{\delta T}{T} = 3\sqrt{\frac{8\epsilon J_k(k\ek)}{3k^2}},
\end{equation}
over one revolution of the pendulum. However, a single such revolution corresponds to many orbits of the binary and, therefore, represents a cumulative effect of order $\sqrt{\epsilon}$. It is noteworthy that our particular $\ham_{\textrm{pert}}$, being of an oscillatory nature, does not induce a net drift in phase space -- if the system starts out in the vicinity of the resonant surface, it will remain there but undergo oscillations about it with a magnitude $\sim \sqrt{\epsilon}$. The pendulum approximation developed in this section is valid at all times, albeit for systems near a resonance solely.
For $k = 1$, $\ek = 1/2$ and $\epsilon \approx 10^{-13}$, we obtain $\delta T \approx 4 \times 10^{-7} ~T$. This change is accumulated over about $10^{6.5}$ periods. This is considerably greater than the $\mathcal{O}(\epsilon)$ change accumulated over a few orbital times (i.e. a time interval much smaller than the libration timescale), but is only visible when observing the binary for a fairly long amount of time (depending on the precision of the observations, of course).

In addition to the narrow libration region of width $\sim \sqrt{\epsilon}$, fast circulation far away from resonances affect the instantaneous position of the binary. This effect is not strictly taken into account in the resonant formalism developed in \cite{Arnold} (but cf. appendix \ref{sec:integrable}), but we can treat it approximately as follows. Suppose the system is given an initial condition $\mathcal{O}(\epsilon^p)$ from the resonant surface, where $0< p < 1/2$. This is still near the resonance. However, \eqref{eq:1} will hold only if the "error" term $\mathcal{O}(\hat{J}_\res^3)$ is negligible relative to the "kinetic" term $\propto \hat J_\text{res}^2$ and the potential term $\propto\epsilon$. The other harmonics in $\ham_\textrm{pert}$ involving both $\beta$ and $\gk$, still average out to zero on a dynamical timescale and, thus, are irrelevant for the long-time evolution of the system. To assess whether $\mathcal{O}(\hat{J}_\res^3)$ can be neglected, notice that, if $\hat{J}_\res = c_1 \epsilon^p$ at $t=0$, then the "kinetic" term dominates over both the "potential" and "error" term since it is of order $\mathcal{O}(\epsilon^{2p}) \gg \mathcal{O}(\epsilon,\epsilon^{3p})$ at $t=0$. If we restrict ourselves to the range $1/3<p<1/2$, then the ``potential'' term is the second dominant term whereas the "error" term is still negligible, so that \eqref{eq:1} is recovered.

Let us pursue the discussion with this additional restriction for simplicity. In this regime, the variation of $\hat{J}_\res^2$ is $\mathcal{O}(\epsilon)$, which in turn implies
\begin{align}
  & \delta\hat{J}_\res \cdot \hat{J}_\res(t=0) \sim \epsilon, \\ &
  \mbox{i.e.}\quad \delta\hat{J}_\res \sim \epsilon^{1-p}.
\end{align}
One is led to the conclusion that, if one plots the variation in $J_c$ -- which is linear in $\hat{J}_\res$ -- as a function of the initial distance from $J_c^*$, this variation (owing to the proximity to the resonant surface) behaves like $\epsilon/\abs{J_c(t=0) - J_c^*}$ (cf. figure \ref{fig2}).

Initial conditions further away than $\mathcal{O}(\epsilon^{1/3})$ from the resonant surface are not captured by this approximation. In reality, $\epsilon$ is extremely small, whence the observed binaries will be most likely off resonance. We, therefore, require the aid of numerical simulations to study the behaviour far from resonance. This is the subject of sec. \ref{sec:nbody} below.
 
 \section{N-body simulations}\label{sec:nbody}

In this Section, we numerically examine the dynamics of binary pulsar systems under the influence of scalar field oscillations. We consider axion masses in the range $10^{-22}-10^{-20}\  \rm eV$ bracketing the limits set thus far by the Lyman-$\alpha$ forest measurements \cite{Armengaud:2017nkf,Irsic:2017yje}. Furthermore, we purposely pick up a large dark matter density of $\rho_{\rm DM} = 5\cdot 10^3\ \rm  M_{\odot}\  pc^{-3}$ to ensure that the simulation results are not significantly affected by numerical noise. For an axion mass $m_a=10^{-21}\ \rm eV$, such a density is admittedly achieved only within the hypothetical, $\sim 1\ \rm pc$ - radius solitonic core residing in the central region of the Milky-Way. Notwithstanding, we emphasize that our choice of $\rho_{\rm DM}$ mainly serves the purpose of illustrating the response of the binary system near and away from resonances.

For our fiducial model, we choose an axion mass $m_a=10^{-21}\  \rm eV$ and a binary mass $M=2M_{\odot}$, so that the resonant values of the semi-major axis satisfy $\ak = 0.205 k^{2/3} \rm AU$. Higher binary masses would physically make more sense since the typical mass of Neutron stars is expected to be larger than the Chandrasekhar mass of $1.4 M_\odot$. Nevertheless, this approximation does not hinder our illustration purposes. Higher binary masses will simply shift the resonances as can be seen from Eq. \eqref{eq: kth resonance}.

To examine the dynamics of the binary pulsar perturbed by the oscillating axion field, we integrate the binary with the extra instantaneous force given in Section \S\ref{sec:secular} using direct N-body simulations. For the N-body integration, we use the publicly available code \texttt{REBOUND} \citep{ReboundMain}.
We use \texttt{IAS15}, a fast, adaptive, high-order integrator for gravitational dynamics, accurate to machine precision over a billion orbits \citep{ReboundIAS15}. Our numerical simulations can capture the impact of the axion oscillations even near resonances, as the system is integrable to first order in the perturbation (see appendix \ref{sec:integrable}).

 \begin{figure}[h!]
 \centering
 	\includegraphics[width=10cm]{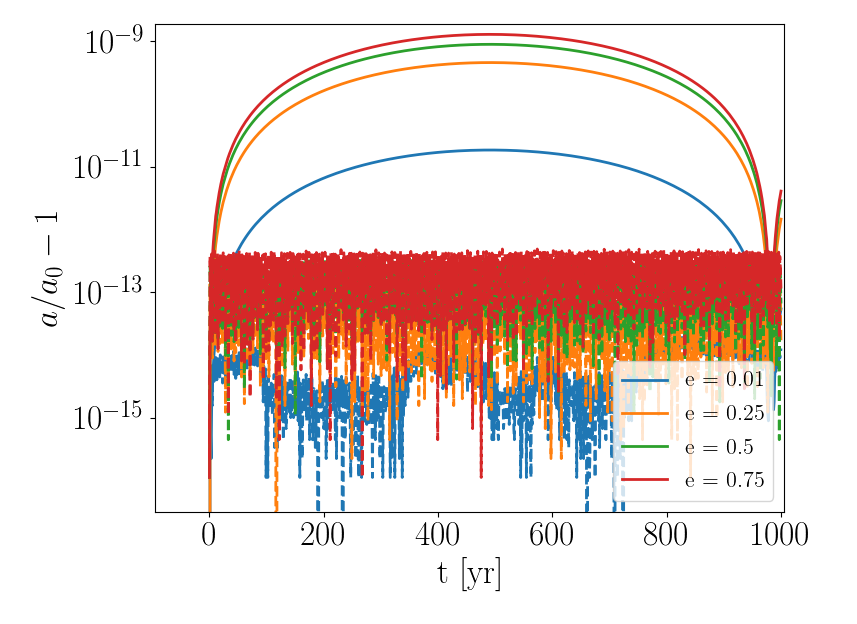}
 	\caption{Example of resonant and non-resonant orbits. The axion mass is $m_{\rm{a}}=10^{-21} \rm eV$. The solid lines show the behaviour of a resonant orbit, with the initial separation is $a_0 =a_*$ where $a_*=0.2052 \ \rm AU$ is the SMA at the fundamental ($k=1$) resonance. The dashed lines show the behaviour of a non-resonant orbit, with the initial separation is $a_0 = 0.9 a_*$ .
 	}
 	\label{fig1}
 \end{figure}

Fig. \ref{fig1} shows the evolution of the SMA as a function of time for resonant and non-resonant orbits with various eccentricities. The resonant orbits with SMA $a_* = 0.2052 \ \rm AU$  accumulate instantaneous changes to achieve a fractional perturbation of order $|a(t)/a_0- 1| \sim 10^{-9}$ after a few hundred periods. They exhibit a periodic pattern of characteristic timescale of $\sim 1000 \ \rm yr$ consistent with the $\sin(\gk+\varphi)$ dependence of the period derivative $\dot{T}$.
By contrast, the non-resonant simulations (dashed) have no long time coherence (i.e. $a/a_0-1$ fluctuates on a very short timescale relative to the resonant orbits) and only lead to a fractional perturbation $|a(t)/a_0- 1|$ of order $\sim 10^{-13}$, four orders of magnitude lower than in the resonant case, and still two orders of magnitude lower when compared to a resonant and nearly circular orbit ($e=0.01$).

Note that, in both cases, the circular orbits remain almost unperturbed, while larger eccentricities result in larger effects, as expected from \cite{Blas}.

In order to explore the structure of the resonances as a function of mass and SMA, we run a grid of initial conditions and plot the maximal change $\Delta a / a_0$, where $\Delta a \equiv (\max(a) - \min(a)) / 2$. We sample three axion masses  $m_a=10^{-20}, 10^{-21}, 10^{-22}\  \rm eV$. For each mass, we initialize the orbit with $2000$ different values of the SMA, with a log-uniform sampling centered on the lowest resonant peaks. This leads to a total of $6000$ simulations. The initial eccentricity is $e=0.5$ and all other angles are set to zero. The total time of integration is $1000$ orbital periods, and is determined separately for each binary.

\begin{figure}[h!]
  \centering
  \includegraphics[width=8.5cm]{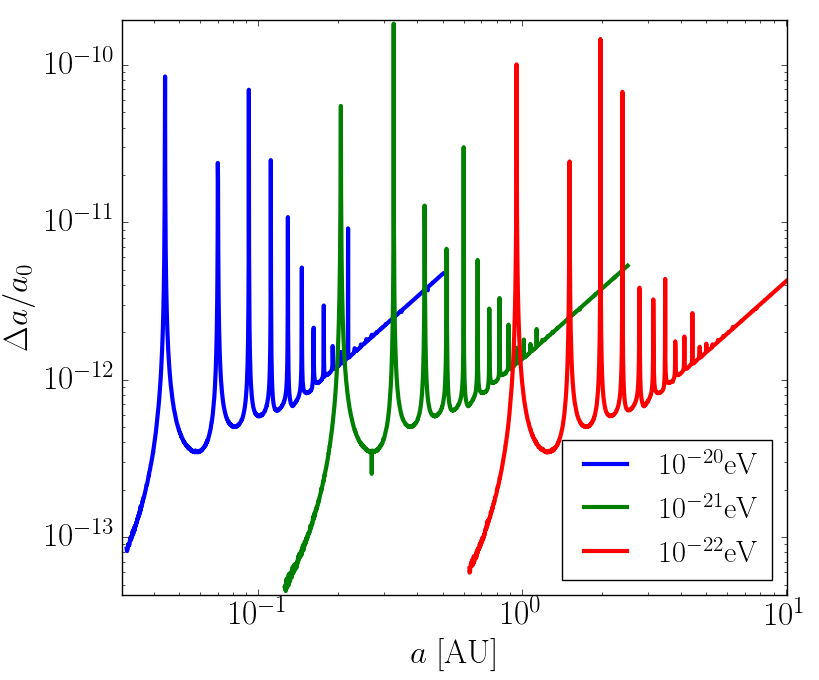}
  \caption{Maximal change of the SMA $\Delta a / a_0 = (\max(a) - \min(a))/2a_0$ as a function of the initial one. Each line corresponds to a different axion mass  ($m_a=10^{-20}, 10^{-21}, 10^{-22}\  \rm eV$ for blue, green and red lines, respectively). The initial eccentricity is $e=0.5$. Each curve exhibits several resonances. The relative width and amplitude reflects a universal behaviour independent of the axion mass. }
  \label{fig2}
 \end{figure}

The results are presented in Fig. \ref{fig2}. The maximal relative variation of the SMA, $\Delta a / a_0$, exhibits a series of resonant peaks, whose width and amplitude hardly changes as the axion mass is varied, except for an overall frequency shift. This shows that the resonant pattern is driven by the ratio $\Omega/\omega_a$.  The structure of those peaks is given by equation \eqref{eqn:resonant change}, when transforming $T$ to $a$ via Kepler's third law. Note also that $\Delta a / a_0$ scales like $\propto a^{3/2}$ when $a\gg a_0$. This reflects the scaling $\dot{T}\propto T^2$ in Eq.(\ref{eq:blastime}), which implies $\dot{a}\propto a^{5/2}$ through Kepler's 3rd law.

In order to compare our findings to the secular change derived in \cite{Blas}, we use the results obtained in Fig. \ref{fig1} in the resonance regime. Namely, we note that the maximal change of $a$ is about $a_{\rm max} \sim a_0(1 + 10^{-9})$. Converting this change into a perturbation of the orbital period, we find  $T_{\rm max} \approx T_0 (1 + 1.5\times 10^{-9})$. As a result, the derivative of the (secular averaged) period at the fundamental resonance ($k=1$) can be estimated as
\begin{equation}
    \dot{T} \approx \frac{T_{\rm max} - T_0}{t_{\rm res}} = 10^{-9}\frac{3T_0}{2t_{\rm res}} \approx 2 \cdot 10^{-13}.
\end{equation}
where, in our case, the unperturbed period is $T_0 \approx 24 \ \rm days$ and the time to reach the maximum resonance is $t_{\rm res} \approx 500\ \rm yr$. Ignoring the slowly varying term $\sin(\gk+\varphi)$ in Eq.~(\ref{eq:blastime}), we estimate the period derivative derived in \citep{Blas} as $\dot{T} = 2.83 \cdot 10^{-13}$.  This is likely an overestimate since the sine modulation $\sin(\gk+\varphi)$ must gradually decrease during the evolution until $t=t_{\rm res}$, so that it vanishes at $T_{\rm max}$. The period derivative reflects this behaviour and also decreases monotonically to zero over the same time interval.

To conclude, our numerical simulations reproduce the secular results of \citep{Blas} at exact resonance with reasonable accuracy. Most importantly, they allow us to explore the instantaneous changes induced by the axion oscillations near and far away from resonances, which was not considered in \cite{Blas}. Our next goal is to calculate the signal-to-noise ratio (SNR) for a detection of these instantaneous changes in the measurements of the time of arrival (TOA) signal of an hypothetical binary pulsar.
We will focus on an axion mass $m_a = 10^{-21}\  \rm eV$ for illustration.

 \section{Prospects for detection with pulsar timing experiments}\label{sec: prospects}

A detailed introduction to the pulsar timing technique can be found in, e.g., \cite{pulsar_book,handbook_pulsar}. The timing method is based on repeated measurements of the TOA of high signal-to-noise average pulses (whose shape is fairly stable). This technique has been used to monitor the TOAs of particularly interesting pulsars over the past decades. The data is further processed using pulsar timing packages, the most popular one being \texttt{TEMPO2} \cite{tempo2}.

The best timing accuracy achieved thus far by the Parkes radio telescope is approximately tens of $\mu$s \citep{2003MNRAS.340.1359F} for 100-1000 of individual observations. The Square Kilometer Array (SKA \cite{2009IEEEP..97.1482D}) might achieve a timing accuracy as high as $5$ ns, while routinely decrease the uncertainty in the TOA down to values of order 10 -- 100~ns for a large number of MSPs \citep{2011A&A...528A.108S}.

Various non-relativistic and relativistic effects contribute to fluctuations in the TOA. The Newtonian contribution includes the R{\o}mer time delay, while the relativistic corrections include the Einstein and Shapiro time delay \citep{handbook_pulsar}. Since we have explored the impact of axion oscillations in the Newtonian regime, we will focus on the R{\o}mer time delay. A similar analysis can be carried out for relativistic pulsars using Post-Newtonian corrections to the Newtonian dynamics. This analysis is out of the scope of the current paper and should be performed elsewhere.

 \subsection{Signal-to-noise for the R{\o}mer delay}\label{sec: romer}

 We analyze the typical pattern of the TOA of the average pulses of a typical binary in and off resonance and discuss the SNR and the possibility to constrain the mass and density of axion dark matter cores.

 The square of the SNR for a detection of axion coherent oscillations in the TOA measurements is
 \begin{equation}
 \left(\frac{S}{N}\right)^2 = \frac{1}{\sigma_\Delta^2}\sum_{i=1}^N \big[\Delta t_\text{\tiny TOA}(t_i)\big]^2,\label{eq:snr0}
 \end{equation}
 $\Delta t_\text{\tiny TOA}(t_i)=|\delta\vr(t_i)|/c$ (note that we reintroduced the speed of light $c$ here, for the convenience of the calculation) is the difference in TOA induced by the
 axion oscillations, $|\delta\boldsymbol{r}(t_i)|$ is the difference at time $t_i$ between the perturbed and unperturbed orbits (staring with the same initial conditions at time $t=0$), $t_j=j\Delta$ are the times at which the observations are performed, $N=t_{\rm obs}/\Delta$ is the total number of measurements and $\sigma_\Delta$ is the error on the measurement. The latter can be as small as $\sigma_\Delta=10^{-6}$ s when the pulse shape is averaged over a time interval $\Delta=10$ s. In practice however, TOA measurements are performed only during a fraction of the total observational time. Therefore, we shall make the conservative choice $\sigma_\Delta=10^{-6}$ s and $\Delta=10^4$ s which corresponds to a total precision of $\sigma_\Delta / \sqrt{N}\approx 5$~ns in our calculation of the SNR. On the one hand, this precision is comparable to that expected to be reached by the SKA on a short timescale (forty hours) for average brightness MSPs. On the other hand, it is rather conservative, given the total observation time of 10 years. We choose this accuracy because MSPs discovered in the Galactic centre region are expected to be dimmer than the known MSP population, so it should take more time to achieve a similar timing precision.

 \begin{figure}
   \centering
   \includegraphics[width=8.5cm]{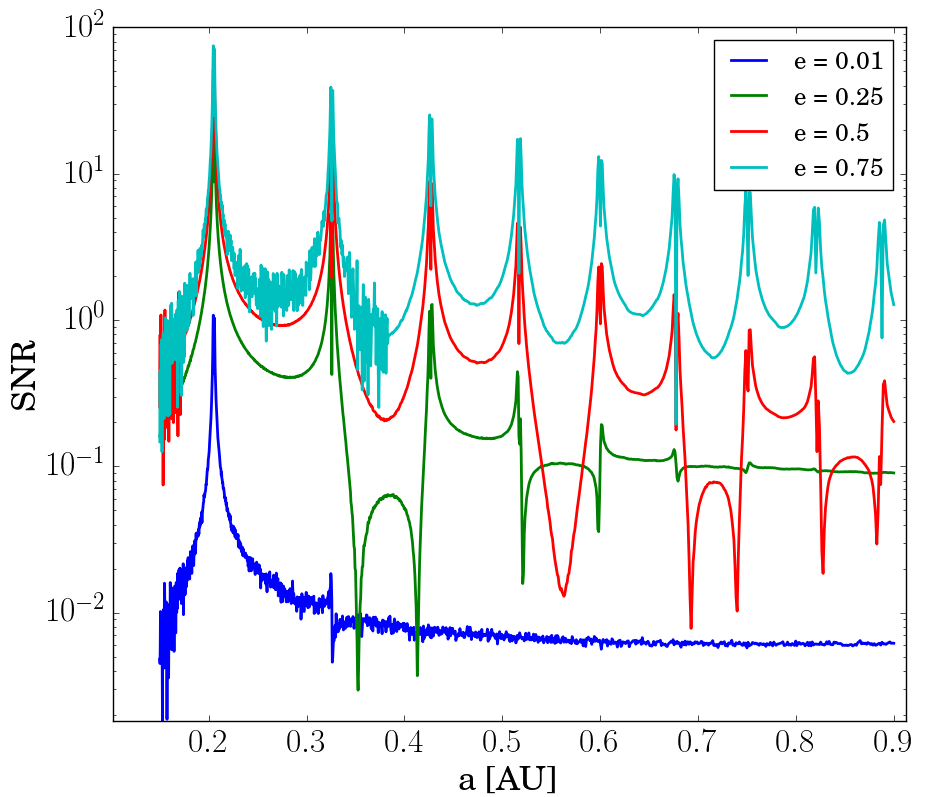}
   \caption{Example of TOA signal-to-noise ratios for various eccentricities. The initial masses are $m_1 = m_2 = M_{\odot}$. The axion mass is $m_{\rm{a}}=10^{-21} \rm eV$, while the DM density is $\rho_{\rm DM} = 5\cdot 10^3\ \rm  M_{\odot}\  pc^{-3}$. The integration time is $t_{\rm end} = 10 \ \rm yr$, with measurements every $\Delta = 10^4\ \rm s$, corresponding to $N=31557$. The measurement error is $\sigma_{\Delta} = 10^{-6}\ \rm s$.  The signal is stronger for more eccentric orbits, and higher order resonances are evident.}
   \label{fig3}
 \end{figure}

Fig.~\ref{fig3} shows the obtained signal to noise ratio pattern of binary pulsar as a function of their initial semi-major axis end eccentricity. For each eccentricity, we sample 1200 different values of the semi-major axis from a log-uniform grid in the range $a_0 / \rm AU \in [0.15, 0.9]$. We integrate each orbit twice, with and without the perturbations and record the difference in the positions $\delta \vr(t_i)$ between the orbits, and plot the SNR defined in equation \eqref{eq:snr0}. Furthermore, we assume a DM density of $\rho_{\rm DM} = 5\cdot 10^3\ \rm  M_{\odot}\  pc^{-3}$, which corresponds to the central density of an axion core in a Milky-Way size halo \cite{Chavanis:2011zi}. This large density ensures that our measurements are well above the numerical noise. The overall amplitude of the SNR scales with $\epsilon\propto \rho_\text{DM}$.

The SNR displayed in Fig. \ref{fig3} show, as expected, that the more eccentric binary pulsars ($e\gtrsim 0.75 $) can be detected more easily, since higher orders of resonance are present and the overall SNR level away from the resonant peaks is fairly high.
Intuitively, we might expect  this behaviour since the instantaneous orbital velocity of more eccentric binary 'samples' a wider range and could be sensitive to several resonances even away from resonance.
By contrast, resonances in systems with lower eccentricities $e\lesssim 0.3$ become harder to detect.
Moreover, as discussed in \cite{Blas} and as we can conclude from considerations of conservation of angular momentum, (nearly) circular orbits are not significantly affected by resonances. It is important to stress that binary pulsars would most likely be found away from the resonant peaks.
Overall, the ratio between the SNR for a system at resonance, and the SNR for a system at the mid-point between two adjacent resonances, can vary considerably. It is around two orders of magnitude for the lowest resonance, and generally decreases for higher order resonances.

Finally, let us also mention that, for SMA $a\gg a_0$ corresponding to much higher orders of resonance (not shown in Fig.\ref{fig3}), we checked that the SNR behaves like $\propto a^{5/2}$, in agreement with the secular estimate of \cite{Blas}. However, the average SNR departs significantly from this scaling around the lowest order resonances shown in Fig.~\ref{fig3}.

 \subsection{Signal-to-noise as a function of axion mass}\label{sec:SNR}

 \begin{figure}
   \centering
   \includegraphics[width=10cm]{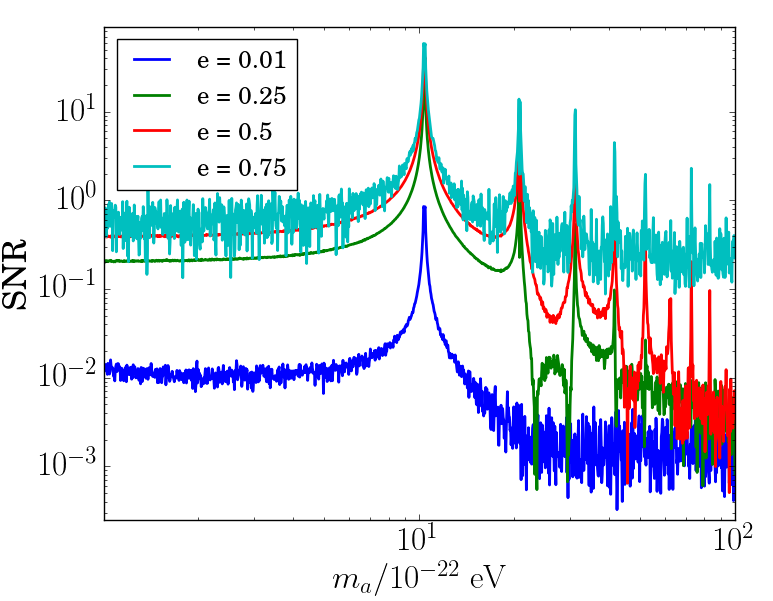}
   \caption{Same as Fig. \ref{fig3}, but with a constant SMA $a_0=0.2\ \rm AU$ and different axion masses as labelled in the figure. The DM density is fixed to a value of $\rho_{\rm DM} = 5\cdot 10^3\ \rm  M_{\odot}\  pc^{-3}$.}
   \label{fig4}
 \end{figure}

In Fig. \ref{fig4}, the SNR is shown as a function of the axion mass for a given dark matter density $\rho_{\rm DM} = 5\cdot 10^3\ \rm M_{\odot}\  pc^{-3}$, and a few binary pulsar systems with identical (unperturbed) SMA but different eccentricities. As a result, $\epsilon$ is constant in these simulations. Therefore, the plateau at low values of $m_a$ indicates that the SNR is proportional to $\epsilon$ times a function which asymptotes to a constant for $m_a\ll \Omega$. The situation is less clear on the high frequency side of the fundamental reasonance, also because we have not pushed our simulations to axion mass values $m_a\gg\Omega$.

For the largest eccentricity, our large dark matter density yields a SNR of order unity away from the resonances. In the central region of our galaxy, $\rho_{\rm DM} \sim 1 \ \rm  M_{\odot}\  pc^{-3}$ (assuming there is no axion core), which translates into a SNR of order $10^{-4}$. If binary pulsar systems are found within $\sim 0.1 {\rm Kpc}$ of the galactic center, then the required sensitivity of SNR$\sim 1$ could be achieved with a twofold improvement on the measurement error $\sigma_\Delta$ and a very dense sampling of the TAO (i.e. a very short time interval $\Delta$ between measurements). Note that our analysis does not take into account correlations in the TOA of different binary pulsars \cite[see, e.g.][for related work]{Martino2018}

 \section{Summary and conclusions}\label{sec:conclusions}

 In this paper, we analyzed the motion of binary pulsars perturbed by an ultra-light axion oscillating background. We used numerical integrations to track the dynamics of the binary and its features. While we recovered the secular result of \citep{Blas} near resonance, we emphasized that the near resonance orbits undergo a libration around a (stable) fixed point which is, strictly speaking, one of the only two possible orbits at exact resonance. These arise when the resonance condition is satisfied, and when the oscillatory force induced by the axion condensate is maximal (in absolute value) at the periapsis.
In addition, we explored in details the short-term, instantaneous features near and far away from resonance which were not considered in previous literature. Although we adopt the ultra-light axions model throughout, our results straightforwardly generalize to other (possibly interacting) models of Bose-Einstein condensate (BEC) dark matter.

While the short-term effects and subsequent time-of-arrival measurements are stronger near the narrow resonances, especially for eccentric binaries, they are also apparent away from the resonant peaks. To quantify their imprint in real data, we considered time-of-arrival (TOA) measurements and computed the signal-to-noise (SNR) ratio required to observe this effects by current pulsar timing techniques. A detection - or lack - of anomalies in the TOA of binary pulsars would constrain the axion mass. The SNR can vary considerably depending on whether one sits near a resonance, or away from them. For the largest value of the eccentricity considered here ($e=0.75$), we find that the broadband SNR level remains approximately constant (resonances excluded) across two orders of magnitude in axion mass.
Though current observations could probe these effects away from resonances only at densities of $\rho_{\rm DM} \approx 10^4 \ \rm M_{\odot} \ pc^{-3}$, future instruments could reduce the sensitivity by at least one or two orders of magnitude. This does not take into account the possibility of cross-correlating TOA measurements from different binary systems.

\acknowledgments

We would like to thank Hagai B. Perets for helpful discussions. E.G. and Y.B.G. acknowledge support from the Technion  Jacobs scholarship; V.D. acknowledges support by the Israel Science Foundation (grant no. 1395/16).


\appendix

 \section{The resonant bit of the perturbation}
 \label{app:resonancebit}
 
We work in the extended phase space $(\boldsymbol{\theta},\mathbf{J})$, where $\boldsymbol{\theta}$ denotes the four angles $(\theta_a,\theta_b,\theta_c,t)$ and $\mathbf{J}$ the four actions $(J_a,J_b,J_c,-E)$. The angle-action variables $(\theta_a,\theta_b,\theta_c,J_a,J_b,J_c)$ are the Delaunay elements of the orbit. In our simplified planar setup, $(\theta_a,J_a)$ are trivially conserved. Hamilton equations read
\be
\dot{\boldsymbol{\theta}}=\frac{\partial\ham}{\partial\mathbf{J}}\;,\qquad \dot{\mathbf{J}}=-\frac{\partial\ham}{\partial\boldsymbol{\theta}},
\ee
where a dot designate a derivative w.r.t. the affine variable $\tau$ introduced to parametrize the phase space trajectory of the system. Note that $\dot{t}=1$ since the extended Hamiltonian $\ham$ is constructed such that it depends on $E$ only through an additive term $-E$. As a result, $\dot{\ham}=0$ along any orbit of the extended phase space.

The $r^2$-dependence of the perturbation Hamiltonian can be expanded into a Fourier series.
As a result, $\ham_\text{pert}$ depends on the actions $J_b$ and $J_c$, but only on the $2\pi$-periodic angle $\theta_c$  ($\theta_b$ is cyclic, so that $J_b$ is a constant of motion).
The smallness of $\epsilon$ and the time-periodic nature of $\ham_\text{pert}$ guarantee that the actual motion remains close to the unperturbed Keplerian orbit.
 
More precisely, since $\ham_\text{pert}$ is an even function of $r^2$, the coefficients of the sine terms all come out zero upon integrating over one (unperturbed) orbit, and we are left with the cosine series
\begin{equation}
  \frac{r^2}{\ak^2} - 1 = \frac{1}{2}\alpha_0(e) + \sum_{n=1}^{\infty}\alpha_n(e)\cos(n\theta_c) \;.
\end{equation}
Here, $\ak$ is the semi-major axis at the $k$-th resonance, see Eq.(\ref{eq: kth resonance}). 
An inner product with $\cos(k\theta_c)$ gives
\begin{equation}\label{eqn:fourier cos coefficients}
  \alpha_n(e) = \frac{1}{\pi}\int_{0}^{2\pi}\left(\frac{r^2}{\ak^2} - 1\right)\cos(n\theta_c)d\theta_c.
\end{equation}
Substitution of the eccentric anomaly $\eta$ turns this expression into
\begin{align}
  \alpha_0(e) &= 3 e^2 \\
  \alpha_n(e) &= -\frac{4}{n^2}J_n(ne) \quad (n\geq 1) \nonumber \;.
\end{align}
Therefore
\begin{equation}
  \frac{r^2}{\ak^2} = 1 +\frac{3}{2}e^2 - \sum_{n=1}^{\infty}\frac{4}{n^2}J_n(ne)\cos(n\theta_c),
\end{equation}
where $J_n(x)$ are Bessel functions.
In order to identify the resonant piece of the Hamiltonian, we use the trigonometric identity
\begin{align}
  \cos(\omega_a t &+ \varphi)\cos(n\theta_c)
  = \frac{1}{2}\big[\cos(\omega_a t + n\theta_c + \varphi) + \cos(\omega_a t - n\theta_c + \varphi)\big] \nonumber \;.
\end{align}
Here, $\varphi$ is the phase of the axion field relative to the unperturbed orbit (whose periapsis is assumed to occur at $t=0$).
We assume that both $\omega_a$ and $\Omega_c = \dot{\theta_c}$  are positive (the same works if one is negative, or both) in the unperturbed system
described by the Hamiltonian
\begin{equation}
  \label{eq:UnpertHam}
  \ham_0 = -\frac{G^2M^2\mu^3}{2J_c^2}-E \;.
\end{equation}
The resonance condition \eqref{eq:resonancecondition} thus reads $\omega_a = k\Omega_c({\bf J}^*)$.
An asterisk will hereafter designate quantities evaluated at the resonant surface.
This defines the resonant surface $J_c=J_c^*\equiv (kG^2 M^2\mu/\omega_a)^{1/3}$,
which is a line in the plane $(J_c,E)$. Picking up the zero energy level (which is arbitrary) further selects a point $(J_c^*,E^*)$ on this surface.

To understand how $J_c$ and $E$ evolves near resonance, it is enough to consider the resonant piece of the perturbation Hamiltonian,
\begin{equation}
  \ham_\text{pert}\supset\ham_\text{res} =-\frac{2\epsilon}{k^2}J_k(k\ek)
  \cos(\omega_a t-k\theta_c+\varphi)\frac{GM\mu}{\ak}\;,
\end{equation}
where $e\approx\ek$ is the eccentricity at resonance.
The difference $\ham_\text{pert}-\ham_\text{res}$ oscillates on a short (dynamical) timescale and, therefore, can be neglected.
Therefore, we write the perturbed Hamiltonian $\ham=\ham_0+\ham_\text{pert}$ (in the extended phase space) near the $k$-th resonance as
\begin{align}
  \label{eq:approxHam}
  \ham(\theta_c,J_c,t,-E) &\approx -\frac{G^2M^2}{2J_c^2}-E +\ham_\text{res} \\
  & = -\frac{G^2 M^2\mu^3}{2 J_c^2} - E - \frac{2\epsilon}{k^2} J_k(k\ek)\cos(\omega_a t-k\theta_c+\varphi) \frac{GM\mu}{a_*} \nonumber \;.
\end{align}
The resulting equations of motion for the actions are
\begin{align}
  \dot{E} &= \frac{\partial\ham}{\partial t} = \omega_a \frac{2\epsilon}{k^2}J_k(k\ek) \sin(\omega_a t-k\theta_c+\varphi) \frac{GM\mu}{a_*} \\
  \dot{J}_c &= - \frac{\partial\ham}{\partial \theta_c} = \frac{2\epsilon}{k}J_k(k\ek) \sin(\omega_a t-k\theta_c+\varphi) \frac{GM\mu}{a_*} \nonumber \;.
\end{align}
The physical interpretation of these equations is straightfoward: $\dot{E}$ represents the rate at which energy is injected into the system (averaged over
one cycle as we ignore the fast oscillations), whereas  $\dot{J}_c$ describes the variation of the usual energy constructed from the conservative part of
the potential (i.e. the 2-body gravitational interaction).
Conservation of energy ensures that they are proportional to each other, i.e. $\frac{\dot{E}}{\omega_a}=\frac{\dot{J}_c}{k}$. 
The relation
\begin{equation}
  \dot{T}= \frac{6\pi}{\mu^3}\left(\frac{J_c}{GM}\right)^2 \dot{J}_c\;,
\end{equation}
along with the second Hamilton's equation
$\dot{J}_c=-\partial\ham_\res/\partial\theta_c$, leads to the equation \eqref{eq:blastime} derived in \cite{Blas}.

Consider now an orbit initially on the resonant surface, i.e. $(J_c(0),E(0))=(J_c^*,E^*)$.
The resonant condition implies that, at leading order in $\epsilon$, $|E|$ and $|J_c|$ grow linearly with time. As a result, the orbit will gradually leave the
resonance surface. However, the argument of the $\sin$ factor also starts evolving, and this will bring the orbit back towards the resonant surface.
Since the derivatives $\dot{E}$ and $\dot{J}_c$ are in phase, the actual motion is an oscillation around $(J_c^*,E^*)$ on a line at an angle $\arctan(k/\omega_a)$
with the resonant surface, i.e. $\frac{E-E^*}{\omega_a}=\frac{J_c-J_c^*}{k}$.

In order to reveal the oscillations around the resonant surface, it is convenient to work with (non-)resonant canonical variables.
We define the (non-)resonant angle $\gk$ (resp. $\beta$) through the canonical transformation
\begin{equation}
  \label{eq:ResAngles}
  \left(\begin{array}{c}
    \gk \\
    \beta
  \end{array}\right) = \left(\begin{array}{cc}
    \omega_a & -k \\
    \omega_a & +k
  \end{array}\right)\left(\begin{array}{c}
    t \\
    \theta_c
  \end{array}\right).
\end{equation}
The corresponding actions are
\begin{equation}
  \label{eq:ResActions}
  \left(\begin{array}{c}
    J_\textrm{res} \\
    J_\textrm{nr}
  \end{array}\right) = \left(\begin{array}{c}
    -\frac{E}{2\omega_a} - \frac{J_c}{2k} \\
    -\frac{E}{2\omega_a} + \frac{J_c}{2k}
  \end{array}\right).
\end{equation}
The new set of variables $(\gk,\beta,J_\text{res},J_\text{nr})$ captures the near resonance orbits defined by the Hamiltonian \eqref{eq:approxHam}, which have
$J_\text{nr}=J_\text{nr}^*=0$ and $J_\text{res}$ oscillating around $J_\text{res}^*$. Only the orbits with $\gk+\varphi=0$ or $\pm\pi$ never leave the resonant surface.
They have $J_\text{res}=J_\text{res}^*$, and correspond to a stable ($\gk+\varphi=\pm\pi$) and an unstable ($\gk+\varphi=0$) fixed point in the Poincar\'e section
$(\beta,J_\text{nr})=(0,0)$ of the phase space.
 
To proceed further, we recast the resonant part of the perturbation Hamiltonian into the form
\begin{equation}
  \ham_\text{res}(\gk) =-\frac{2\epsilon}{k^2}J_k(k\ek) \cos(\gk+\varphi)\frac{GM\mu}{\ak}\;,
\end{equation}
and use \eqref{eq:ResActions} to express the actions $J_c$ and $-E$ around the resonant surface as
\begin{align}
  J_c &= J_c^*-k\left(J_\text{nr}^*-J_\text{nr}\right)-k\left(J_\text{res}-J_\text{res}^*\right)\approx J_c^* -k\left(J_\text{res}-J_\text{res}^*\right) \\
  -E &= -E^*+\omega_a\left(J_\text{nr}-J_\text{nr}^*\right)+\omega_a\left(J_\text{res}-J_\text{res}^*\right)\approx -E^*+\omega_a\left(J_\text{res}-J_\text{res}^*\right) \nonumber .
\end{align}
The final approximations arise, again, from the fact that the non-resonant actions oscillate rapidly around their resonant value. On defining 
$\hat{J}_\text{res}=J_\text{res}-J_\text{res}^*$, the previous relations allow us to expand the unperturbed part \eqref{eq:UnpertHam} of the Hamiltonian around $J_c^*$ and $-E^*$.
Since
\be
\frac{1}{J_c^2}=\frac{1}{J_c^{*2}}\left[1+2\frac{k\hat{J}_\text{res}}{J_c^*}+3\left(\frac{k\hat{J}_\text{res}}{J_c^*}\right)^2+\dots\right]\;,
\ee
the resonant condition implies that the term linear in $\hat{J}_\text{res}$ vanishes.
This leads to equation \eqref{eq:1}, which shows that the {\it near resonance orbits} are described by a simple pendulum Hamiltonian.
The only trajectories that are at exact resonance - the {\it resonant orbits} - correspond to the fixed points mentioned above.

To conclude this Section, we note that the fast oscillations can be treated - in a first approximation - as stochastic fluctuations around the smooth phase space
trajectories described by \eqref{eq:1}.

 \section{An Integral of Motion}
 \label{sec:integrable}

Three integrals of motion are trivial to find -- those are the two independent components of the angular momentum, and the extended phase space Hamiltonian.  To further investigate the motion of the binary under the perturbation in equation \eqref{eqn:axion force}, let us try to find an integral of motion that takes into account all the resonances.

The method is described in \cite{LichtenbergLieberman}. We consider a function $I(\boldsymbol{\theta},\mathbf{J}) = I_0(J_c) + \eps I_1(\boldsymbol{\theta},\mathbf{J})$. 
For it to be an invariant, it needs to have vanishing Poisson bracket with the Hamiltonian, that is, $\{I,\ham\} = 0$. This implies that, in the cosine expansion of appendix \ref{app:resonancebit},
\begin{equation}
  \label{eqn:definition of invariant}
  (\omega_a + n\Omega_c)I_n = nA_n\frac{dI_0}{dJ_c},
\end{equation}
where $A_n$ comes from the Fourier expansion of $\ham_\textrm{pert} = \sum_{n=-\infty}^{\infty}\eps A_n\cos(n\theta_c + \omega_a\tau)$, related to $a_n(e)$ via equations \eqref{eqn:definition of epsilon} and \eqref{eqn:fourier cos coefficients}. We set $\varphi=0$ without restricting the validity of our argument.

The important part is that the left-hand-side of equation \eqref{eqn:definition of invariant} vanishes at the resonances, so we take $\frac{dI_0}{dJ_c}$ to be a function that vanishes there, too. We choose
\begin{equation}
  \frac{dI_0}{dJ_c} = \sin\!\left(\frac{\pi\omega_a J_c^3}{G^2M^2}\right),
\end{equation}
whence (up to an arbitrary constant)
\begin{equation}
  \label{eq:invariant}
  I  = \frac{(GM)^{2/3}}{3(\pi\omega_a)^{2/3}}\,\mathrm{Si}\!\left(\frac{1}{3},\frac{\pi\omega_a J_c^3}{G^2M^2}\right)  + \eps\sum_{n=-\infty}^{\infty}\sin\!\left(\frac{\pi\omega_a J_c^3}{G^2M^2}\right)\frac{nA_n\cos(n\theta_c + \omega_a\tau)}{n\Omega_c(J_c) + \omega_a} \;,
\end{equation}
where $\textrm{Si}(a,z) =\int_0^z t^{a-1}\sin tdt$. $I$ is a constant of motion to first order in $\eps$; together with the angular momentum components $J_a, J_b$ and the extended phase-space Hamiltonian, we have $4$ integrals of motion, and therefore the system is integrable to first order in $\eps$; the system is restricted to move along phase space trajectories with $I = \textrm{const},~ \ham = \textrm{const}$. The total variation in $J_c$ may therefore be estimated by taking the variation of equation \eqref{eq:invariant}, and using the implicit function theorem.

\label{lastpage}

\bibliography{refs}

\end{document}

%% file: defs.tex
\def\lsim{~\rlap{$<$}{\lower 1.0ex\hbox{$\sim$}}}
\def\bsim{~\rlap{$>$}{\lower 1.0ex\hbox{$\sim$}}}

\def\apjl{Astrophys.\ J.\ Lett.}
\def\mnras{Mon.\ Not.\ R.\ Astron.\ Soc.}
\def\nat{Nature (London)}

\def\aap{Astron.\ Astrophys.}
\def\apj{Astrophys.\ J.}

\def\apjs{Astrophys.\ J. Supp.}

\def\prd{Phys.\ Rev.\ D}

\def\physrep{Phys. Rep.}

\def\mathbi#1{\textbf{\em #1}}

\def\vr{\mathbi{r}}